\documentclass[12pt,english,review,authoryear]{elsarticle}
\usepackage[T1]{fontenc}
\usepackage[latin9]{inputenc}
\usepackage{geometry}
\usepackage{float}
\usepackage{units}
\usepackage{amsbsy}
\usepackage{graphicx}
\usepackage{color}
\usepackage{natbib}

\makeatletter
\@ifundefined{date}{}{\date{}}
\usepackage{babel}

\@ifundefined{showcaptionsetup}{}{%
 \PassOptionsToPackage{caption=false}{subfig}}
\usepackage{subfig}
\makeatother

\usepackage{babel}
\begin{document}

\title{Fracture of granular materials composed of arbitrary grain shapes: a new cohesive interaction model}

\author[]{A. Neveu \corref{cor1}}
\ead{aurelien.neveu@ifsttar.fr}
\author[]{R. Artoni}
\author[]{P. Richard}
\author[]{Y. Descantes}

\address{LUNAM Université, IFSTTAR, MAST, GPEM, F-44340 Bouguenais, France}

\cortext[cor1]{Corresponding author}

\begin{abstract}

Discrete Element Methods (DEM) are a useful tool to model the fracture of cohesive granular materials. For this kind of application, simple particle shapes (discs in $2D$, spheres in $3D$) are usually employed. However, dealing with more general particle shapes allows to account for the natural heterogeneity of grains inside real materials. We present a discrete model allowing to mimic cohesion between contacting or non-contacting  particles whatever their shape in $2D$ and $3D$. The cohesive interactions are made of cohesion points placed on interacting particles, with the aim of representing a cohesive phase lying between the grains. Contact situations are solved according to unilateral contact and Coulomb friction laws. In order to test the developed model, $2D$ unixial compression simulations are performed. Numerical results show the ability of the model to mimic the macroscopic behavior of an aggregate grain subject to axial compression, as well as fracture initiation and propagation. A study of the influence of model and sample parameters provides important information on the ability of the model to reproduce various behaviors.

\end{abstract}


\maketitle

\section{Introduction}

Studying the crushing of cohesive materials is of importance for a
wide range of natural and industrial processes. In the production
of aggregates, rock blocks are crushed and the resulting fragments
are required to meet high standards mainly in terms of size and shape. Successive
crushing steps are usually carried out to achieve the requested aggregate
characteristics, leading to a waste of good quality raw materials
and a high energy cost that could both be mitigated upon improving
the crushing efficiency. 

To better understand the crushing mechanics
of this kind of material, previous studies have attempted to reproduce
numerically the behavior of aggregates with the aim of linking the
heterogeneous microstructure of the material to its macroscopic behavior
\citep{Cheng2008,Bolton2004,Brown2014,Jing2003,McDowell1998,Spettl2015,Affes2012a}.
Continuous methods like finite element methods (FEM) are not well suited to deal with the complex heterogeneous microstructure of aggregates, due to the computational cost required to properly describe
microscale behavior. The Lattice element method (LEM) stands as a compromise using a network of $1D$ elements 
connected at nodes which are positionned on a regular or irregular lattice, the former being able to carry properties (elastic stiffness, strength) to mimic the behaviour of the different phases of the material (particle, cohesive matrix...). This method has been used to study the fracture of cemented aggregates \citep{Affes2012a}, and multiphase particulate materials \citep{bolander2005,asahina2011,berton2006}.

Numerical simulations using discrete Element Methods (DEM) have also been successfully used to describe the elastic behavior and rupture
mechanism of a rock piece \citep{Potyondy2004,Weerasekara2013a,Bolton2004,Cheng2003,Cheng2008,Jiang2006,Andre2012}.
In these methods, a grain is represented by a collection of particles
with contact bonds to model cohesion inside the material. In order
to describe materials such as concrete, in which grains are surrounded
by a cementitious matrix, \citet{Hentz2004,Hentz2004a} have introduced an interaction range which mimics cohesion between
two particles even when not in contact. They have shown that increasing this interaction range allows to take into account the degree of interlocking of particles in rock \citep{Scholtes2013}. In most cases, only simple
particle shapes (discs in $2D$, spheres in $3D$) were used because of
the increasing complexity of contact detection and force computation
in the case of more irregular shapes. However, in order to introduce
the natural complexity of grains composing an aggregate, some authors
have chosen to use more irregular particles (polygons), built from
a Voronoï tessellation \citep{D'Addetta2002,Galindo-Torres2012,Nguyen2015},
or by clustering spherical particles \citep{Cho2007,Zhao2015}. Furthermore,
using polygons in $2D$ or polyhedra in $3D$ allows to build samples
with a high solid fraction up to $1$. 

In this paper, we introduce a cohesive interaction model for Discrete
Element Methods. It allows to model cohesion between contacting and
non contacting particles, and suits any kind of particle shape. Since
the aforementioned model deals independently with cohesive interactions
and solid contacts, it can be used in the framework of non-smooth
or smooth methods. In the following, section $2$ first describes
the method used to solve contact situations and then gives a comprehensive
description of the cohesive interaction model which has been developed.
Section $3$ presents results obtained with $2D$ uniaxial compression
tests and illustrates the influence of the model parameters. Finally,
section $4$ is devoted to the conclusions and the perspectives of
this work.

\section{Contacts and cohesive interactions model}

We aim to develop a model which is applicable to both non contacting
and contacting particles, in order to represent a cemented material composed of particles surrounded by a cohesive phase. This type of material is present both in industrial and natural configurations (concrete, sandstone, ...). It is quite natural to represent this material as a collection of particles interacting through contact and cohesive interactions. We assume that the cohesive behavior is only due to the cement paste, and has a different nature than the contact between particles.
We prefer such a representation with respect to a lattice one to highlight the effect of contact between particles on the modeled material. 

So a mixed method combining the
contact description of rigid particles with the elastic behavior of
cohesive interactions has been developed.

\subsection{Contact between particles}

The contact problem between two particles is solved using the Non
Smooth Contact Dynamics (NSCD) initially developed by \citet{moreau1988}
and \citet{Jean1999}. This method allows to solve long lasting
contacts or collision situations in rigid grains assemblies through
a mechanically based approach that fulfills the non-interpenetration
requirement without assuming regularized laws at contact points \citep{moreau1988}.
The NSCD method is implemented in the code LMGC90 \citep{Renouf2004},
which includes contacts detection between polygons (or discs) in
$2D$ or polyhedra (or spheres) in $3D$.

\subsection{Cohesive interaction}\label{Cohesive_interaction}

Cohesion is set inside the modeled material by applying forces which
oppose relative motion between particles. This relative displacement
is computed between two cohesion points, each placed on one of the
two interacting particles. These points and associated forces will
be denoted as a ``cohesive interaction'' in the following. The cohesive
paste between two particles is thus modeled by one or more cohesive
interactions, i.e. one or more pair of cohesion points. As cohesion
is treated separately from geometrical contacts, this allows to apply
cohesive forces even if particles are not touching each other, whatever
their shape, with the aim of describing a cohesive paste between them.

In the present work, the cohesive paste is represented by two cohesive
interactions in $2D$ and three in $3D$. For a $2D$ system, the
cohesion points are located on both sides of the particles at a distance
$R$ from the center (Fig. \ref{fig:sch_pt_2D}). This allows to resist
relative rotation of the particles as it will induce relative displacements
of cohesion points, and thus a reaction torque which will oppose this
rotation. For a $3D$ system, at least $3$ cohesive interactions
are required for each pair of particles to resist tension/compression,
shearing, twisting and bending (Fig. \ref{fig:sch_pt_3D}). The cohesion
points are thus regularly placed on each interacting particle at the same
distance $R$ from their center, all being located in a plane orthogonal
to the line joining the centers of the two particles. The intensity
of the resistance to rotation can thus be adjusted by changing the
distance between the cohesion points and the center of the particle.

\begin{center}
\begin{figure}[htvp]
\begin{centering}
\begin{minipage}[t]{0.49\columnwidth}%
\begin{center}
\subfloat[\label{fig:sch_pt_2D}]{\begin{centering}
\includegraphics[scale=0.5]{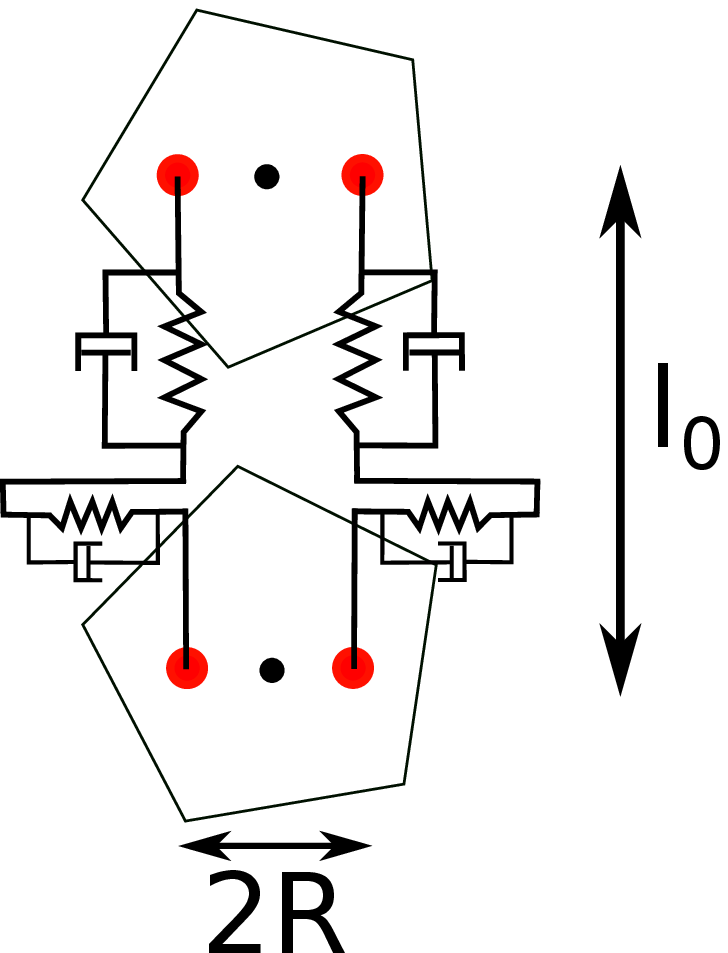} 
\par\end{centering}

}
\par\end{center}%
\end{minipage}%
\begin{minipage}[t]{0.49\columnwidth}%
\begin{center}
\subfloat[\label{fig:sch_pt_3D}]{\begin{centering}
\includegraphics[scale=0.4]{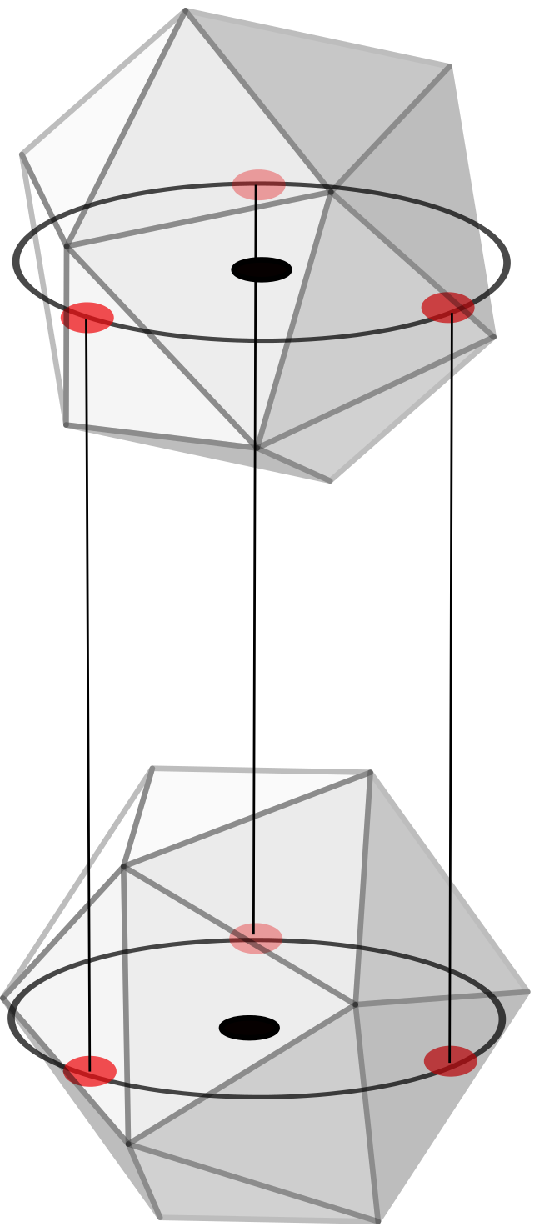}
\par\end{centering}

}
\par\end{center}%
\end{minipage}
\par\end{centering}

\caption{Setting of cohesive interactions between two particles for a $2D$ system
(a) and a $3D$ system (b). Each cohesive interaction is composed of
a spring and a damper in compression/tension and shear.\label{fig:sch_pts}}
\end{figure}

\par\end{center}

\subsubsection{Reaction forces}

The cohesive paste between particles is described as an elastic material
which can be stretched, compressed and sheared. A cohesive interaction
plane, defined for each cohesive interaction, corresponds to the tangential
plane normal to the line of unit vector $\boldsymbol{n}$ which connects
the two cohesion points. The tangential unit vector $\boldsymbol{t}$
is defined by the tangential deformation. The cohesive reaction forces
are computed at cohesion points, and then applied to the center of
mass of the particle. Their expression in the reference frame of the
cohesive interaction plane is: 
\[
\boldsymbol{F}=F_{n}\cdot\boldsymbol{n}+F_{t}\cdot\boldsymbol{t},
\]
where the normal and the tangential components are respectively given
by:

\begin{equation}
F_{n}=-k_{n}^{s}\Delta_{n},\label{eq:fn}
\end{equation}

\begin{equation}
F_{t}=-k_{t}^{s}\Delta_{t}.\label{eq:ft}
\end{equation}
with $\Delta_{n}$ and $\Delta_{t}$ denoting the compressive/tensile
components of the relative displacements in the cohesive interaction
plane reference frame, and $k_{n}^{s}$ and $k_{t}^{s}$ the corresponding
normal and tangential stiffnesses respectively.

This linear spring model does not dissipate energy. Indeed, in real
materials kinetic energy is dissipated by microscopic processes during
contact. Therefore we add dissipation by means of viscous damping.
The normal and tangential reaction forces become:

\[
F_{n}=-k_{n}^{s}\Delta_{n}-k_{n}^{v}U_{n},
\]

\[
F_{t}=-k_{t}^{s}\Delta_{t}-k_{t}^{v}U_{t},
\]
where $U_{n}$, $U_{t}$ denote the relative normal and tangential
velocities and $k_{n}^{v}$, $k_{t}^{v}$ the corresponding damping
coefficients in normal and tangential directions.

It should be observed that, while tension is opposed solely by cohesive interaction forces, compression may be opposed by both cohesive interaction and contact forces, thus ensuring a transition from elastic to perfectly rigid behaviour in the latter case.

\subsubsection{Model parameters}

In order to determine the microscopic coefficients $k_{n}^{s}$ and
$k_{t}^{s}$, an analogy is done with the continuum elastic modulus
of a cohesive paste. Thus, the normal spring stiffness $k_{n}^{s}$
is derived from the Young's modulus $E_{\mu}$ of a deformable cohesive
paste as : 
\begin{equation}
k_{n}^{s}=\frac{E_{\mu}S}{l_{0}},\label{eq:kn}
\end{equation}
where $S$ is the cohesive interaction surface, which can be viewed
as the sectional area of the cohesive paste between the grains divided
by the number of cohesive interactions used to model this paste ($2$
in $2D$, $3$ in $3D$). The initial length of a cohesive contact
$l_{0}$ is the distance between the two cohesion points when the
cohesive interactions are set. Since the cohesive interaction surface
$S$ and length $l_{0}$ differ for each pair of particles, a microscopic
stiffness heterogeneity is thus introduced. The tangential
spring stiffness is determined from the shear modulus, which corresponds to the shear stress to
shear strain ratio. By analogy, the force exerted by the tangential
spring is assigned a value which balances the force needed to shear
the cohesive bond over the tangential displacement $\Delta_{t}$.
Hence, the shear modulus may be expressed as:

\begin{equation}
G=\frac{|F_{t}|l_{0}}{S.\Delta_{t}}.\label{eq:G}
\end{equation}

Replacing in Eq. (\ref{eq:G}) $F_{t}$ by its expression given by
Eq. (\ref{eq:ft}) yields the following expression: 
\begin{equation}
k_{t}^{s}=\frac{GS}{l_{0}}.\label{eq:kst}
\end{equation}

Assuming a perfectly isotropic cohesive paste material, the shear
modulus relates to the Young's modulus $E_{\mu}$ via the Poisson's
ratio $\nu_{\mu}$ according to: 
\begin{equation}
G=\frac{E_{\mu}}{2(1+\nu_{\mu})}.\label{eq:GE}
\end{equation}

Using Eqs. (\ref{eq:kn}), (\ref{eq:kst}) and (\ref{eq:GE}),
the ratio of the normal to tangential stiffnesses may therefore be
expressed as follows: 
\begin{equation}
\frac{k_{n}^{s}}{k_{t}^{s}}=2(1+\nu_{\mu}).
\end{equation}

The damping coefficients $k_{n}^{v}$ and $k_{t}^{v}$ are
calculated to obtain a critical damping which prevents oscillations
at contact. It has to be noted that the stiffnesses of cohesive interactions are computed once and remain constant until breakage. Thus, damage occurs in the sample only by breakage of cohesive interactions. More sophisticated
models such as Cohesive Zone Models (CZM) which take into account
the progressive damage of a cohesive contact can be found in \citep{Raous1999,Riviere2015}.

\subsubsection{Breakage of cohesive interactions}

We assume that in a cemented material, fracture occurs due to the
breakage of the cohesive paste between grains, which corresponds to
the rupture of cohesive interactions. Several bond
rupture criteria for the DEM modeling of cohesive materials are available
in the literature \citep{Jiang2014,Delenne2004}. In the present work,
a cohesive interaction can break only due to tension and/or shear
load. When it breaks, the cohesive interaction is no longer able to oppose relative displacement, which is equivalent to setting its stiffnesses ($k_{n}$, $k_{t}$) to zero.
However, the two (or three in $3D$) cohesive interactions used to model cohesion between a pair of particles are treated independently : if one breaks, the other one remains intact if not above rupture limit. If all the cohesive interactions between two particles are broken, only contact interaction remains active. Note also that breakage is not reversible in this model : a broken cohesive interaction cannot be restored.
 Besides, the model incorporates a threshold displacement above which rupture of the cohesion interaction occurs for both tension and shear. For tension, Eqs. (\ref{eq:fn}) and (\ref{eq:kn}) yield:

\begin{equation}
\Delta_{n}^{max}=\frac{l_{0}}{E_{\mu}}\,.\,\sigma_{r}^{n},\label{eq:delta_n_max}
\end{equation}

and for shear Eqs. (\ref{eq:ft}), (\ref{eq:kst}) and (\ref{eq:GE}) yield:

\begin{equation}
\Delta_{t}^{max}=\frac{2(1+\nu_{\mu})\,l_{0}}{E_{\mu}}\,.\,\sigma_{r}^{t},\label{eq:delta_t_max}
\end{equation}

with $\sigma_{r}^{n}$ and $\sigma_{r}^{t}$ the strength in tension and shear which are input parameters of the model, taken identical for all the interactions. The threshold displacements have been chosen identical  in tension and shear, which implies from Eqs. (\ref{eq:delta_n_max}) and (\ref{eq:delta_t_max}):

\begin{equation}
\sigma_{r}^{t}=\frac{\sigma_{r}^{n}}{2(1+\nu_{\mu})}\label{eq:sigma_rt}
\end{equation}

In the following, the rupture threshold will be referred as $\sigma_{r}$, with $\sigma_{r}^{n} = \sigma_{r}$ and $\sigma_{r}^{t}$ given by Eq. (\ref{eq:sigma_rt}).

\section{2D Benchmark test}

\subsection{Initial packing}

In the following, the approach described above will be applied to
a $2D$ benchmark, the uniaxial compression of a rectangular sample
composed of polydisperse regular pentagons. The initial packing is
built according to the following method. First, discs of average diameter
$d$ are deposited under gravity in a rectangular box composed of
four rigid walls. The grain diameters are uniformly distributed between
$0.5d$ and $1.5d$ to prevent crystallization. Then, each disc is
replaced by a regular pentagon which fits exactly inside the disc,
with the same arbitrary orientation for all pentagons. The packing is then subjected
to a compression phase to increase its solid fraction. This is done
by moving the walls at a constant velocity ($V=5.10^{-3}\left[d^{\nicefrac{3}{2}}\sqrt{\frac{E_{\mu}}{m}}\right]$,
see section \ref{sub:section_scale} for scales), with no other external
forces than those applied by the walls (i.e. no gravity). The relative
velocities between top/bottom and left/right walls are chosen to achieve
the desired height to width aspect ratio of the packing.

Due to the displacement of the walls, the solid fraction can be higher
close to the walls than in the packing bulk. To achieve homogenization,
the following mixing phase is performed: after a given number of compression
time-steps, walls are kept fixed and random velocities are assigned
to pentagons. To ensure maximum efficiency of the mixing process, the 
pentagon/pentagon and pentagon/wall contacts are solved with a full restitution shock law free of friction, so that kinetic energy is not dissipated.
Then, all velocities are set to zero and the compression phase resumes.

As soon as the target solid fraction is reached, the compression phase
is stopped. All velocities are set to zero and only the locations
of particles are kept. Some precautions have to be taken at walls/granular sample contacts during simulations. Indeed,
due to the sharp shape of particles, the probability of applying the
load to a single particle is high. This could initiate the rupture
of the packing at an early stage of the simulation, in an unrealistic
manner. To avoid this, the pentagons located at the top and bottom
of the packing are smoothed to create a flat wall/packing surface
and so to increase the contact surface area with the walls (Fig. \ref{fig:smooth_polyg}).

\begin{center}
\begin{figure}[htvp]
\begin{centering}
\begin{minipage}[t]{0.49\columnwidth}%
\begin{center}
\subfloat[\label{fig:smooth_polyg}]{\begin{centering}
\includegraphics[scale=0.12]{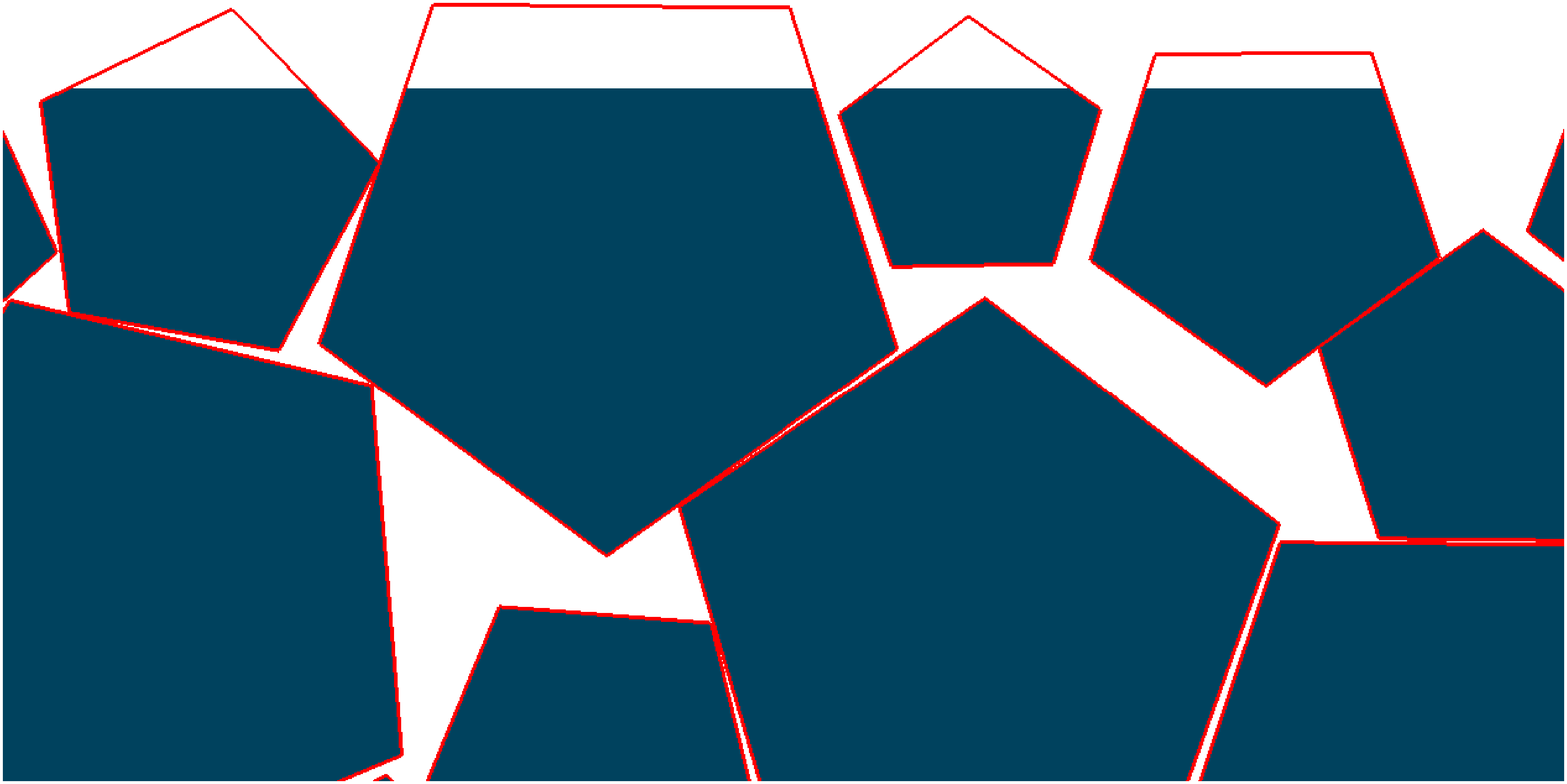}
\par\end{centering}

}
\par\end{center}%
\end{minipage}%
\begin{minipage}[t]{0.49\columnwidth}%
\begin{center}
\subfloat[\label{fig:sample_ini}]{\begin{centering}
\includegraphics[scale=0.12]{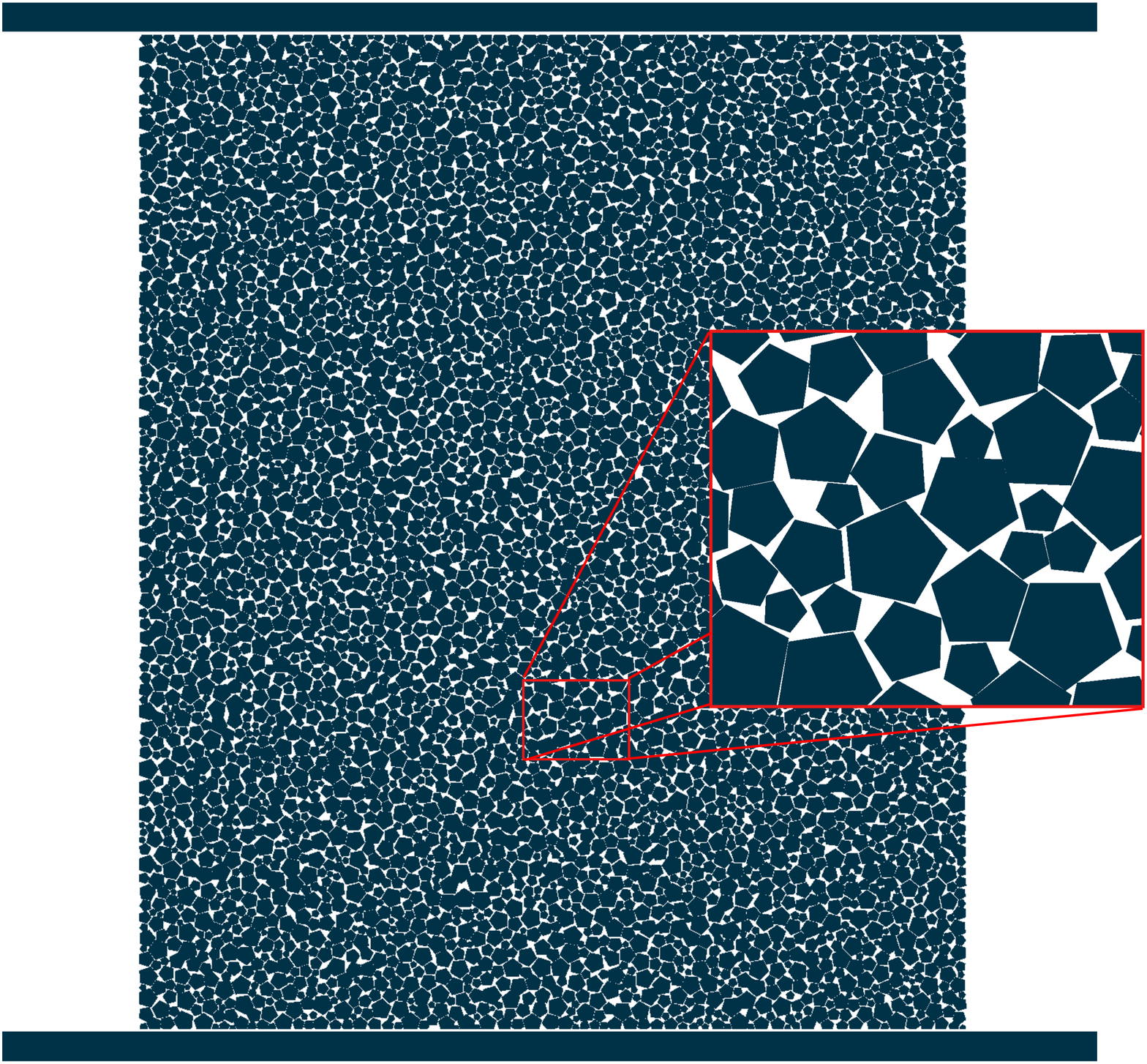}
\par\end{centering}

}
\par\end{center}%
\end{minipage}
\par\end{centering}

\caption{(a) Smoothing of pentagons in the contact boundary with walls. (b)
Typical packing of $5000$ regular pentagons used in compression simulations. }
\end{figure}

\par\end{center}

Note that, as expected, due to the building process which introduces
randomness, two different packings with the same macroscopic properties
(solid fraction, number of particles, size, particle size distribution...)
will generally yield differences in the particle structure. For two
samples, the orientation distributions of normal unit vectors characterizing
contacts as well as cohesive interactions normal unit vector have been plotted in Fig. \ref{fig:distri_orient}. Although
these distributions are similar, small differences are visible between
the samples. We can also note that the orientations are fairly isotropically
distributed. This packing procedure allows to obtain homogeneous samples (in terms of solid fraction, 
random orientations, size distribution) while controlling precisely the final aspect ratio and solid fraction, and thus provides a good benchmark for the model. In addition, this random packing of irregular particles can be representative of some common sedimentary rocks such as sandstone \citep{Yang2013}. The results presented in the following have been obtained from simulations using packings of an aspect ratio of $1.2$ constituted
of $5000$ regular pentagons, which have been found sufficient to allow good reproductability, as well as reasonable computational cost. Such a packing is presented in Fig.
\ref{fig:sample_ini}.

\begin{center}
\begin{figure}[htvp]
\begin{centering}
\begin{minipage}[t]{0.49\columnwidth}%
\subfloat[]{\includegraphics[scale=0.4]{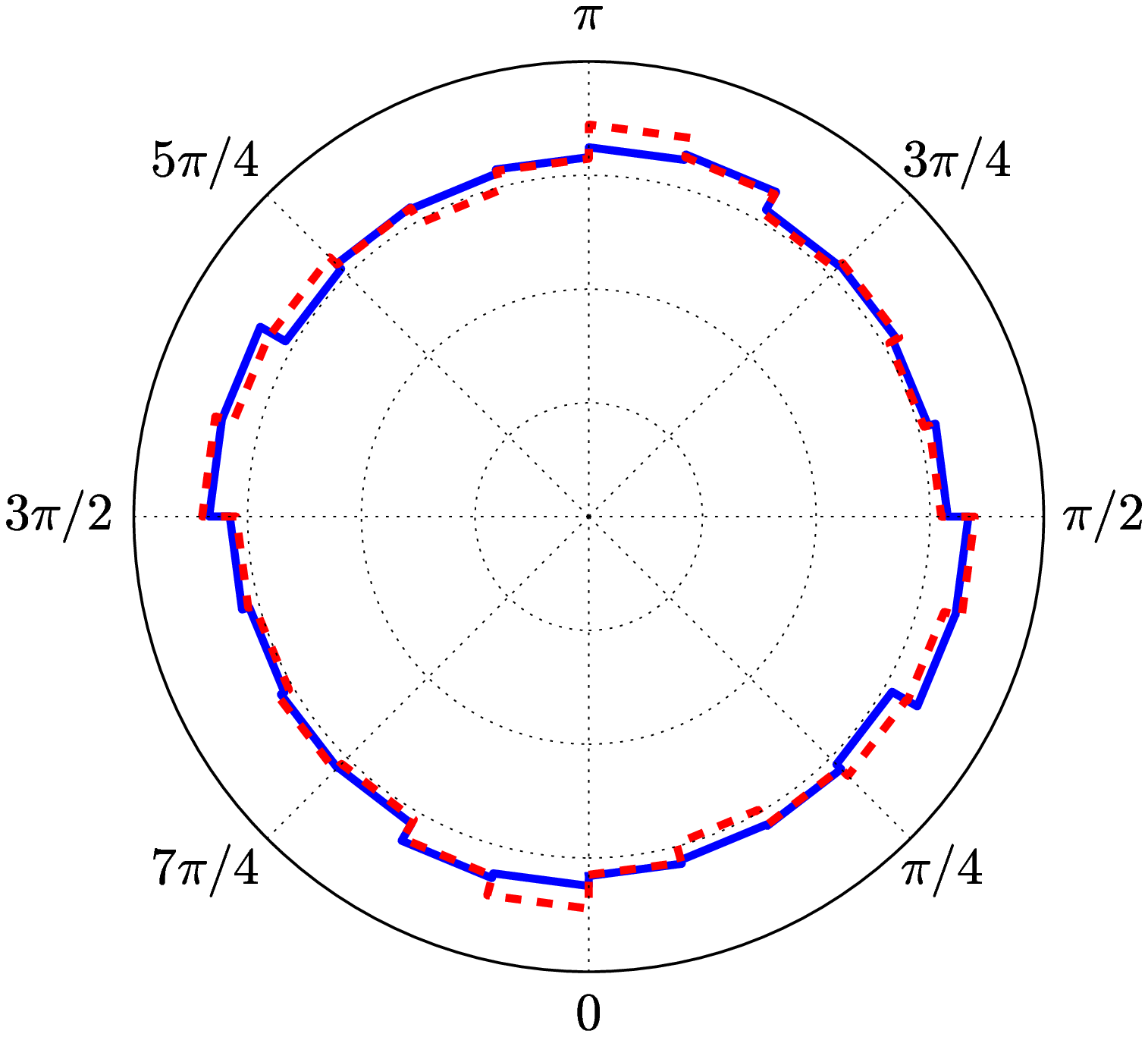}

}%
\end{minipage}%
\begin{minipage}[t]{0.49\columnwidth}%
\subfloat[]{\includegraphics[scale=0.4]{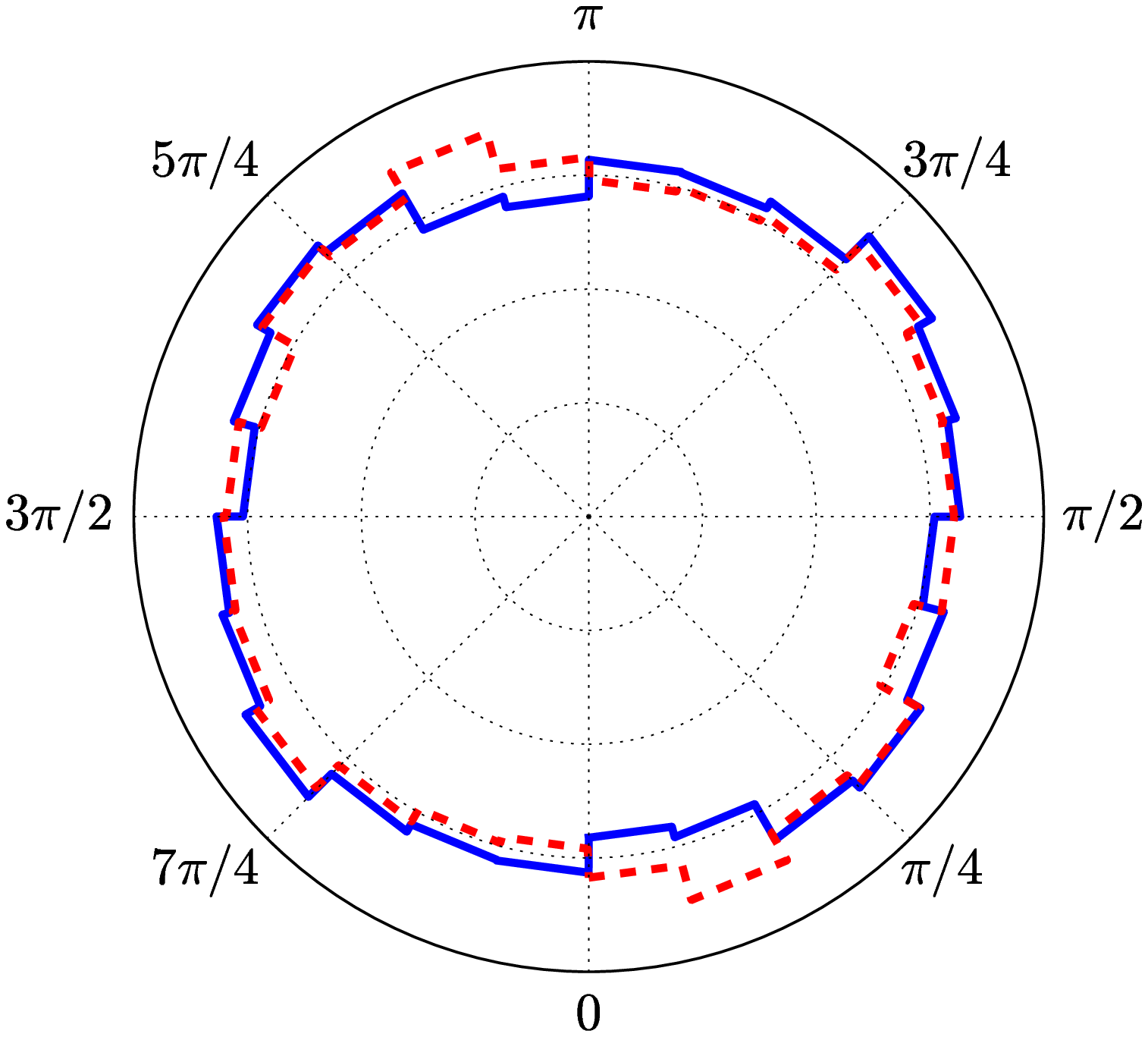}

}%
\end{minipage}
\par\end{centering}

\caption{Orientation distributions of the normal unit vector $\boldsymbol{n}$
for cohesive interactions (a) and contacts (b) obtained for two different
initial samples (plain line, dashed line). \label{fig:distri_orient}}

\end{figure}

\par\end{center}

Once the initial packing is built, the cohesive interactions have
to be set inside the modeled material. As cohesion is active even
if particles are not touching, a criterion is required to decide whether
two particles share a cohesion interaction or not. A rather simple
way to do this is to define an interaction range \citep{Hentz2004}
for cohesive effects, so that if the distance between two particles
is below this range cohesion will act. However, this requires an additional
model parameter, which needs to be added to the calibration procedure.
Another method, which is the one used in this work, consists in determining the list of particles interacting through cohesive bonds by computing a Delaunay triangulation from the centers of all particles. Each edge of the triangulation corresponds to a
pair of particles undergoing cohesive interaction. This choice
is not insignificant because it leads to a constant number of cohesive
interactions per grain \citep{Gervois1995} and thus it may mitigate
the effect of the solid fraction (see discussion in Section \ref{sub:Influence-of-solid}). For each cohesive interaction, two (in $2D$) or three (in $3D$) cohesion points are placed according to the rules defined at the beginning of section \ref{Cohesive_interaction}. The distance $R$ between cohesion points and center of the particles is chosen as the average radius $r$ (the radius of a disc of same area than the pentagon) of the smallest pentagon of the pair of particles enduring a cohesive interaction. A heterogeneity terms of $R$ values  is thus introduced for polydisperse packings, and thus only the $R/r$ ratio is a constant input parameter for all cohesive interactions. A discussion on the influence of $R$ is provided in section \ref{sub:influence_R}.
Once the cohesion is set, the lateral walls are removed and the packing is then ready for loading.

\subsection{Uniaxial Compression}

Uniaxial compression tests have been performed on different packings
by moving a wall at constant velocity $V_{c}$ while maintaining the
other wall fixed. The sample is submitted only to external forces
applied by these walls in the absence of gravity. The compression
velocity is chosen sufficiently low to ensure quasi-static equilibrium
and avoid unexpected material responses such as initiation of rupture
near a wall. The quasi-static equilibrium is checked by comparing
the stresses applied by the two loading walls.

\subsubsection{Simulation parameters\label{sub:section_scale}}

The obtained macroscopic behavior is controlled by microscopic contact
and cohesive interaction parameters. Those used in our simulations
are made dimensionless using the following scaling rules : 
\begin{itemize}
\item Mass scale : mean mass $m$ of particles 
\item Length scale : mean equivalent diameter $d$ of pentagons, defined
as the diameter of a disc with the same surface area as the pentagon 
\item Time scale : $t_{0}=\sqrt{\frac{m}{E_{\mu}d}}$, so that the stress
scale is Young's modulus $E_{\mu}$ of the cohesive paste. 
\end{itemize}
As the Young's modulus $E_{\mu}$ is that of a cohesive paste, it
is taken identical for all cohesive interactions, with $E_{\mu}=1$
according to the chosen stress scale. The Poisson's ratio of
the modeled cohesive paste is taken as $\nu_{\mu}=0.25$. The stiffness
heterogeneity of cohesive interactions originates from geometrical
properties ($S$, $l_{0}$). Friction coefficients at particle/particle
$\mu_{pp}$ and wall/particle $\mu_{pw}$ contacts are set to $0.3$
and $0.5$ respectively. The stress threshold in tension is assigned
a value of $\sigma_{r}^{n}=5.10^{-3}$, and thus following Eq.
(\ref{eq:sigma_rt}) the stress threshold in shear is set to $\sigma_{r}^{t}=2.10^{-3}$.
Finally, the compression velocity is $V_{c}=2.10^{-5}$.

The choice of the time-step $dt$ is important as it guarantees the
precision and the stability of the computation. It has to be sufficiently
small compared to all time scales, including the damping characteristic
time of particles motion, but also not too small to allow reasonable
simulation duration time. A dimensionless time-step value of $10^{-1}$
is considered a good compromise between precision and simulation duration.
Usually simulations require $50000$ to $100000$ time-steps to achieve
rupture of the packing, depending on the value of cohesive interaction
parameters, especially the cohesive interaction strength.

\subsubsection{Macroscopic behavior}

The typical stress/strain curve obtained for uniaxial compression
tests is presented in Fig. \ref{fig:stress-strain-broken} together
with the variations of the number of broken cohesive interactions.
During the first phase of compression, the material clearly exhibits
a linear elastic behavior characterized by the linear part of the
stress-strain curve. During this phase, the cohesive interactions
store the increasing strain energy applied to the packing by the walls.
The modeled packing being initially consolidated, it doesn't exhibit
a non-linear increase of stiffness for small deformations as can be
seen in experiments due to closure of pores or small rearrangements
of grains \citep{wawersik1970}.

Then, the stress supported by some cohesive interactions exceeds the
elastic limit $\sigma_{r}$ and they break. These breakages induce
small drops in the macroscopic stress and the stress-strain curve
starts to deviate slightly form the linear behavior. Suddenly, when
a critical strain $\epsilon_{p}$ is reached, large amounts of cohesive
interactions break: this leads to a major drop of the stress-strain
curve, and the emergence and propagation of a macroscopic crack across
the packing. Although the propagation of the macroscopic crack is
very fast, as we will show, this phenomenon is progressive.

Figure \ref{fig:map} shows the spatial distribution of broken cohesive
interactions at different stages of the compression. Initially, the
cohesive interactions seem to break more or less randomly across the
packing but then just before the peak (II) we observe a localization
and the appearance of a crack path. Moreover, the post-peak region
exhibits several minor peaks which correspond to the extension of the macroscopic
crack or the initiation of a new one. The fracture planes align with
the diagonals of the sample, forming two fracture cones near the walls.
This fracture pattern agrees qualitatively well with experiments of
real cohesive materials under uniaxial compression with frictional
platens \citep{jaeger2007}. However, a different boundary condition (e.g. frictionless
walls) would yield a different fracture pattern \citep{D'Addetta2002}. 

\begin{center}
\begin{figure}[h!]
\begin{centering}
\begin{minipage}[t]{1\columnwidth}%
\begin{center}
\subfloat[\label{fig:stress-strain-broken}]{\begin{centering}
\includegraphics[scale=0.4]{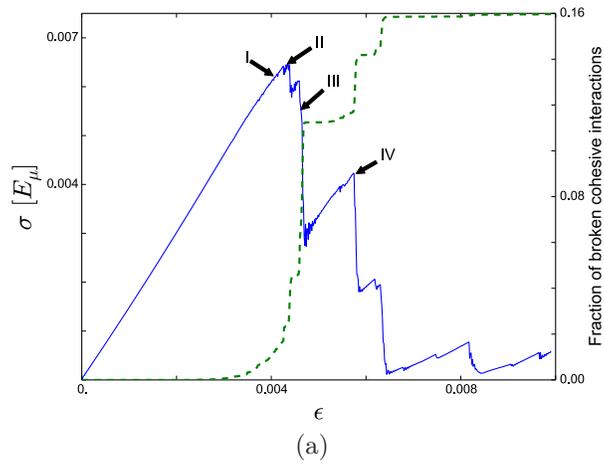} 
\par\end{centering}

}
\par\end{center}%
\end{minipage}
\par\end{centering}

\vphantom{}

\begin{centering}
\begin{minipage}[t]{1\columnwidth}%
\begin{center}
\subfloat[\label{fig:map}]{\begin{centering}
\includegraphics[scale=0.35]{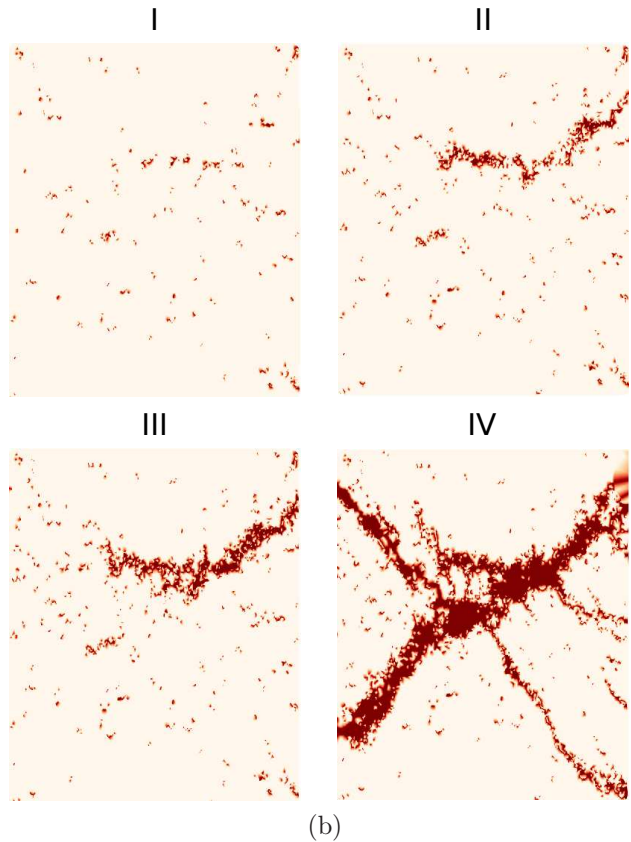}
\par\end{centering}

}
\par\end{center}%
\end{minipage}
\par\end{centering}

\caption{(a) Stress/strain curve (blue solid line) obtained together with the
cumulative share of broken cohesive interactions (green dashed line).
(b) Broken cohesive interactions at different stages of the compression.
The localization of cohesive interactions breakage and progressive
behavior of fracture is clearly seen.}
\end{figure}

\par\end{center}

\subsubsection{Influence of the initial solid fraction\label{sub:Influence-of-solid}}

Different initial packing characteristics yield differences in the
modeled microstructure, and so an expected different macroscopic behavior.
An important packing characteristic is the solid fraction of the sample.
Increasing the solid fraction leads to an increase of the density
of cohesive interactions in the sample, but also to a decrease of
the cohesive interaction length $l_{0}$ since particles are closer
to one another. According to Eq. (\ref{eq:kn}), decreasing $l_{0}$
will increase the cohesive interaction stiffnesses and so influence
the macroscopic stiffness as well. For results presented below, the macroscopic 
Young's modulus is determined as the slope of a fit of the linear part of 
the stress/strain curve corresponding to the linear 
elastic regime. Figure \ref{fig:em_phi} shows 
the macroscopic Young's modulus $E_{M}$ obtained for discrete solid
fraction values ranging between $0.6$ and $0.84$. As expected, increasing the solid fraction leads to an increase of macroscopic Young's modulus, up to $40\%$ in the studied range. It is clear that solid fraction is an important parameter which controls
the macroscopic stiffness of the modeled material. Note also that
as cohesive interactions are set according to a Delaunay triangulation
from the center of particles, the average coordination number of $6$
for $2D$ packings remains the same whatever the solid fraction \citep{Gervois1995}.
Determining the cohesive neighborhood with another method may lead
to a different effect of the solid fraction.

\begin{center}
\begin{figure}[htvp]
\begin{centering}
\begin{minipage}[t]{0.49\columnwidth}%
\begin{center}
\includegraphics[scale=0.4]{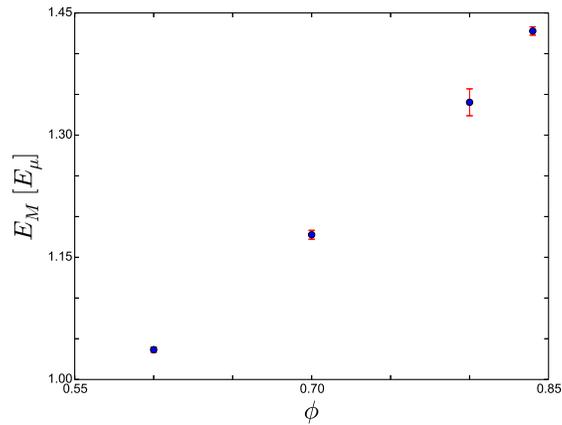} 
\par\end{center}%
\end{minipage}
\par\end{centering}

\caption{Macroscopic Young's modulus $E_{M}$ measured for solid fraction $\phi$
ranging from $0.6$ to $0.84$.\label{fig:em_phi}}
\end{figure}

\par\end{center}

\subsubsection{Influence of cohesive interactions coordination number}

As described before, the Delaunay triangulation used to establish
the cohesive interactions network leads in $2D$ to a cohesive interactions
coordination number $z=6$. In order to study the influence of smaller
$z$ on the macroscopic behavior, cohesive interactions have been
randomly removed from the sample. The stress-strain curves for $z=$
$6$, $5$ ,$4$ and $3$ obtained for samples of solid fraction $\phi=0.84$ are presented in Fig. \ref{fig:stress-strain_crack}.
A linear decrease of $z$ leads to a linear decrease of the macroscopic
Young's modulus because less cohesive interactions contribute to the
overall stiffness of the sample. Therefore, the same strain energy
is endured by fewer cohesive interactions and so the fracture will
occur for a lower stress peak. Moreover, the cohesive interactions
coordination number has a strong influence on the post-peak behavior.
Indeed, the amplitude of the secondary peaks decreases with $z$,
and so does the residual stiffness in the post peak region. This means
that a smaller $z$ leads to a loss of brittleness. The spatial distribution
of broken cohesive interactions at the final stage of the simulation
($\epsilon=0.009$) is plotted in Fig. \ref{fig:map_broken_2}.
For small $z$, it appears clearly that the breakage of cohesive interactions
is no more localized on identified planes. The loss of brittleness
is thus caused by the emergence of a number of small cracks instead
of macroscopic fractures inside the sample.

\begin{center}
\begin{figure}[htvp]
\begin{centering}
\includegraphics[scale=0.4]{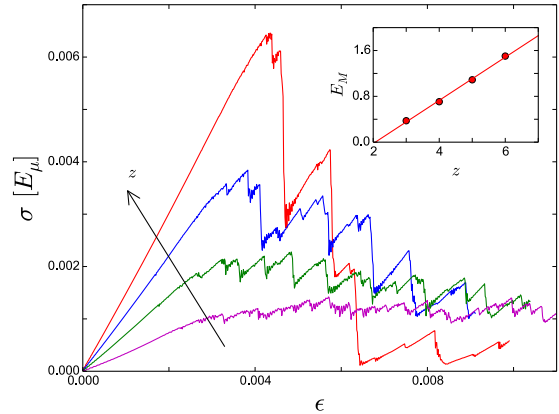}
\par\end{centering}

\caption{Stress/strain curves obtained for values of the cohesive interactions
coordination number $z$ increasing from $3$ to $6$. Insert : macroscopic
Young's modulus $E_{M}$ versus $z$. Error bars are smaller than
the line thickness.\label{fig:stress-strain_crack}}
\end{figure}

\par\end{center}

\begin{center}
\begin{figure}[htvp]
\begin{centering}
\begin{minipage}[t]{0.3\columnwidth}%
\begin{center}
\subfloat[\label{fig:map_broken_2_a}]{\begin{centering}
\includegraphics[scale=0.4]{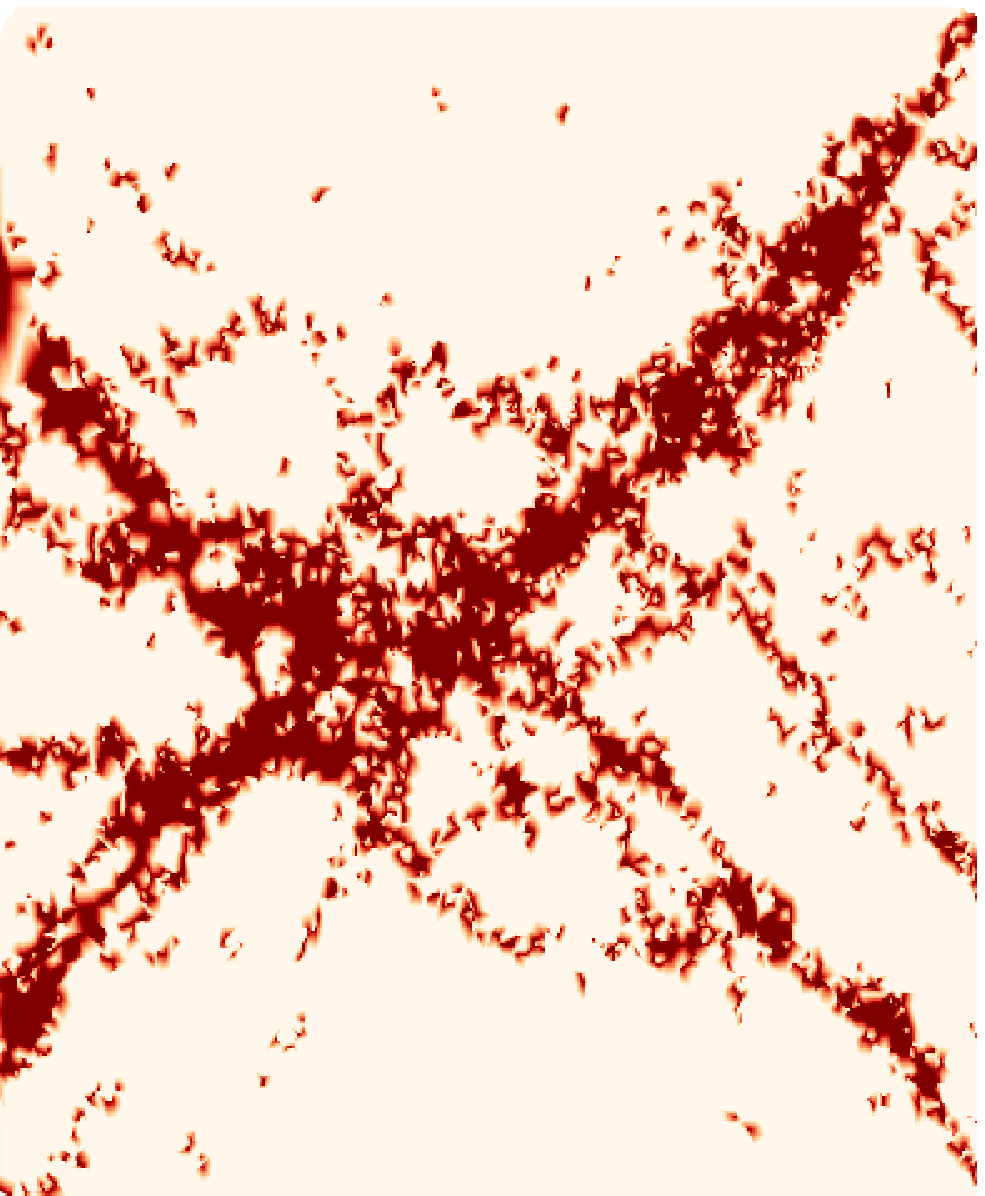}
\par\end{centering}

}
\par\end{center}%
\end{minipage}%
\begin{minipage}[t]{0.3\columnwidth}%
\begin{center}
\subfloat[\label{fig:map_broken_2_b}]{\begin{centering}
\includegraphics[scale=0.4]{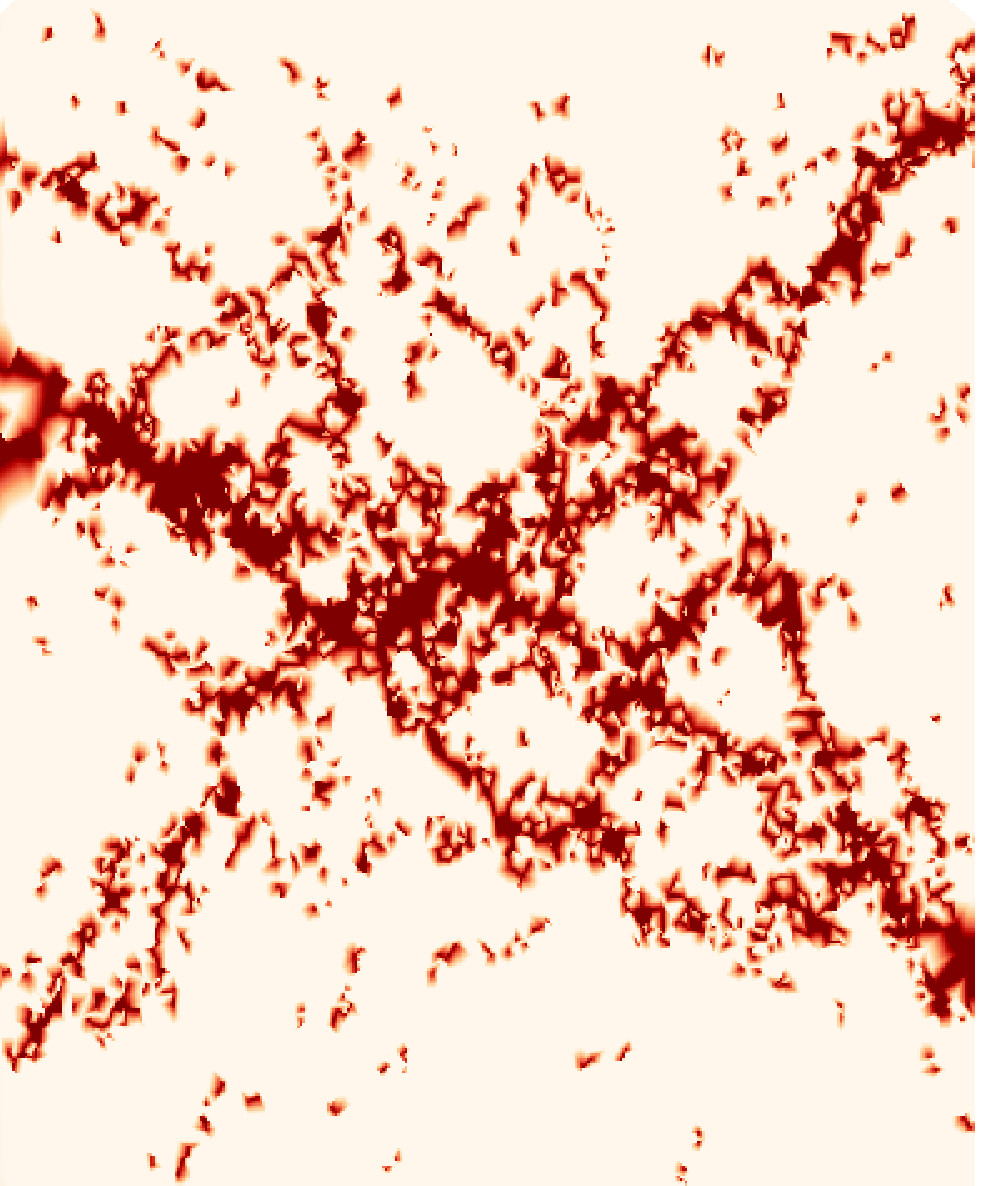}
\par\end{centering}

}
\par\end{center}%
\end{minipage}%
\begin{minipage}[t]{0.3\columnwidth}%
\begin{center}
\subfloat[\label{fig:map_broken_2_c}]{\begin{centering}
\includegraphics[scale=0.4]{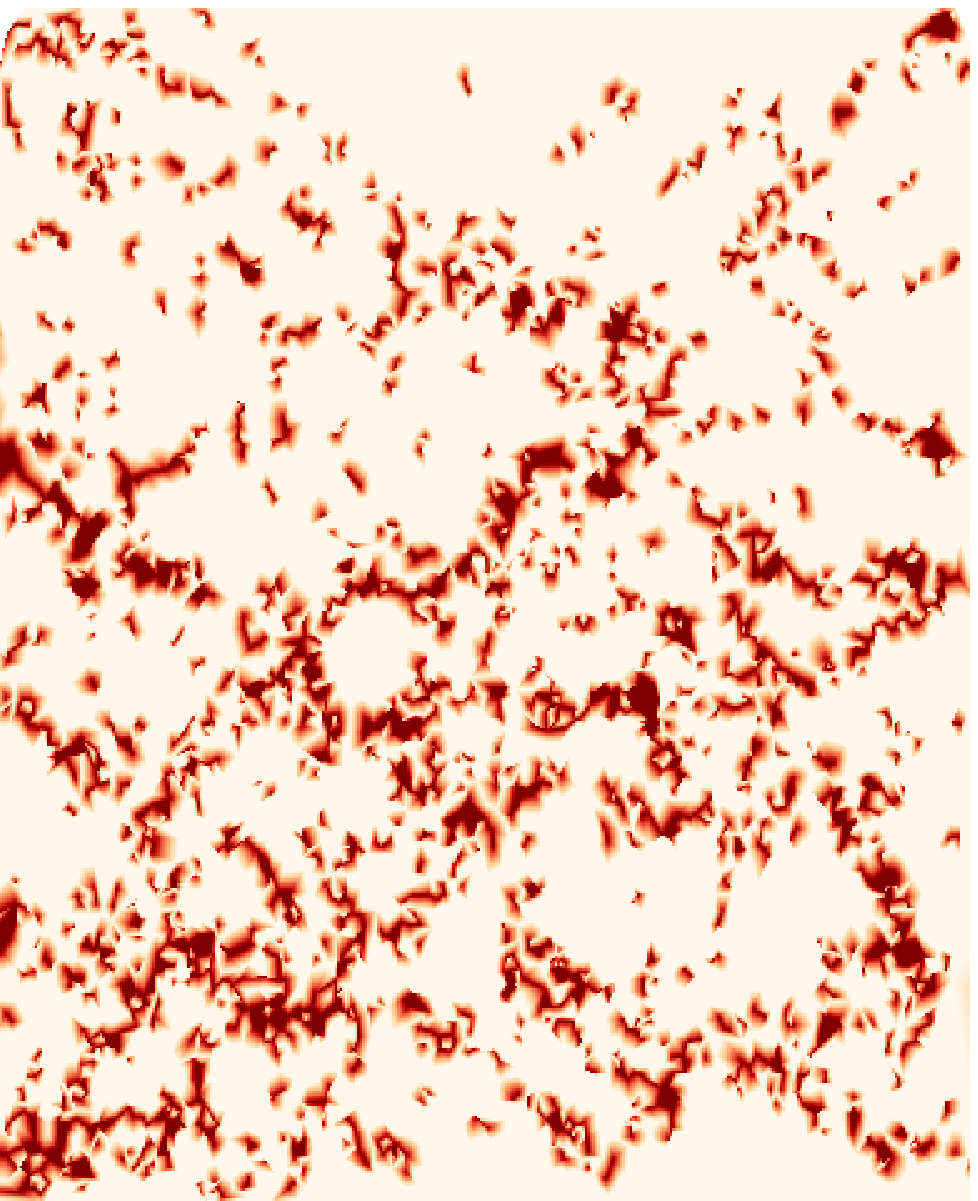}
\par\end{centering}

}
\par\end{center}%
\end{minipage}
\par\end{centering}

\caption{Broken cohesive interactions at $\epsilon=0.009$ for different
cohesive interactions coordination numbers : (a) $z=5$, (b) $z=4$,
(c) $z=3$. A loss of localization for small $z$ is observed. \label{fig:map_broken_2}}

\end{figure}

\par\end{center}

\subsubsection{Influence of contact friction}

In order to evaluate the influence of the particle-particle contact
friction coefficient on the macroscopic Young's modulus $E_{M}$,
several simulations have been performed for values of $\mu_{pp}$
ranging from $0$ to $0.7$ and sample solid fraction values of $0.6$,
$0.7$ and $0.84$ (Fig. \ref{fig:em_mu}). By introducing friction
between particles, reaction forces opposing sliding contribute to
the global resistance of the sample. Increasing the friction coefficient
then yields an increase of the macroscopic Young's modulus characteristic
of the macroscopic stiffness of the material. This effect is stronger
for higher solid fraction values since increasing the solid fraction
also increases the density of particles contacts. However, this effect
is rather small (less than $4\%$) compared to that of the solid fraction.
Indeed, even with a solid fraction of $0.84$, the ratio of the number
of contacts to the number of cohesive interactions is small ($\simeq2\%$), so energy dissipation due to contact friction has a weak influence on macroscopic Young's modulus.

The macroscopic Poisson's ratio $\nu_{M}$, quantifying the dilatational strain under compression of the modeled material, has been determined by considering a centrally located horizontal slice whose thickness is $20\%$ of the sample total height. This slice is then split into $20$ sub-slices of equal thickness, whose mean horizontal strain may easily be calculated. The Poisson's ratio is then determined as the slope of the fit of the vertical strain / horizontal strain curve. Figure \ref{fig:nu_mu} shows the measured $\nu_{M}$ for for different values of the solid fraction and friction coefficient. The values of $\nu_{M}$ vary from $0.145$ to almost $0.2$ in the range of studied parameters. They increase with the solid fraction, which denotes a non negligible effect of contact between particles on their transverse displacement. Indeed, when two pentagons are compressed, the polygonal shape promotes sliding along radial directions, which contribute to dilatancy of the material. The friction coefficient also affects $\nu_{M}$, by opposing lateral displacement via energy dissipation. The Poisson's ratio is thus higher for small values of $\mu_{pp}$. This effect is weaker for lower solid fraction values and seems to saturate for $\mu_{pp}$ greater than $0.3$.

\begin{center}
\begin{figure}[htvp]
\begin{centering}
\begin{minipage}[t]{0.5\columnwidth}%
\begin{center}
\subfloat[\label{fig:em_mu}]{\begin{centering}
\includegraphics[scale=0.4]{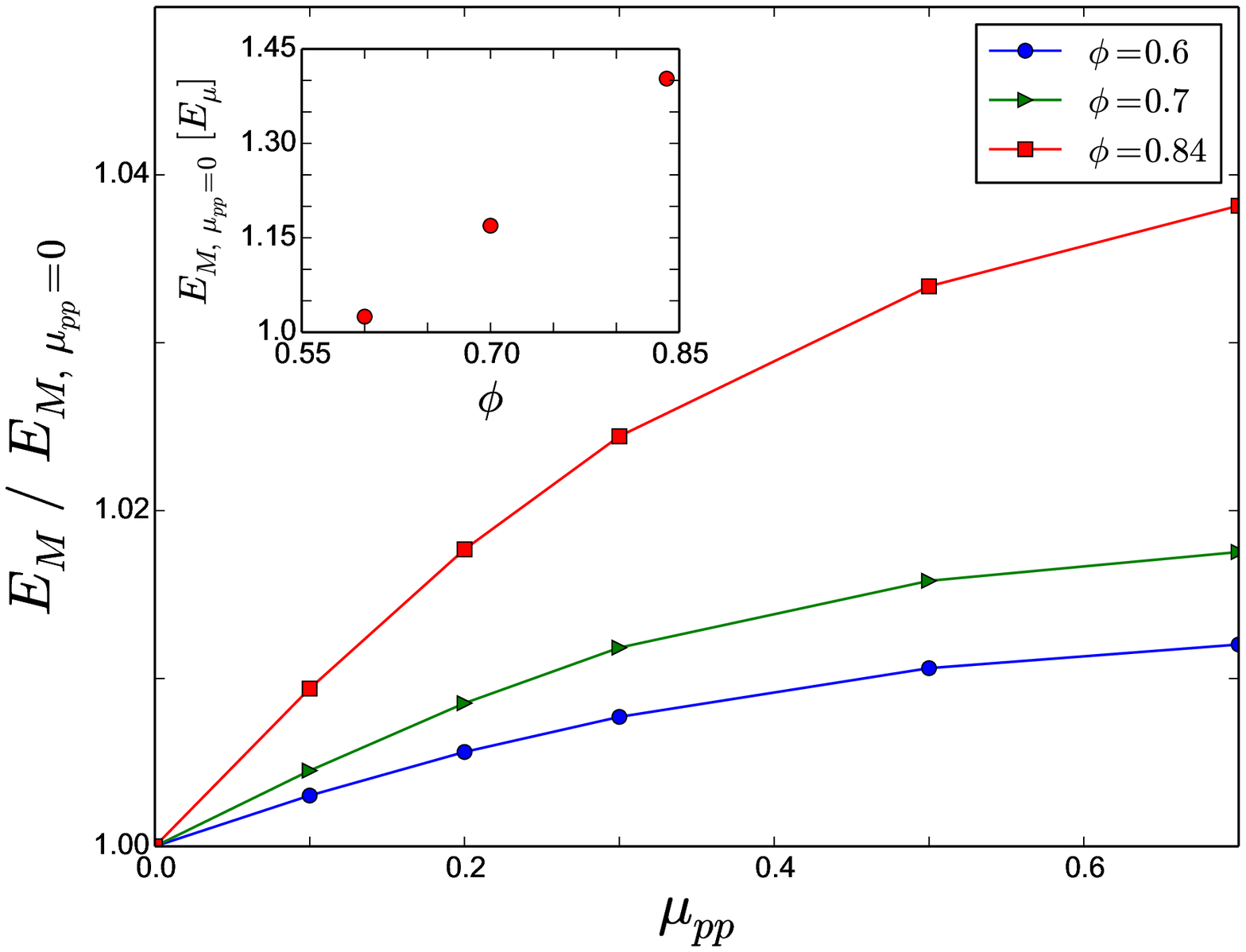}
\par\end{centering}

}
\par\end{center}%
\end{minipage}%
\begin{minipage}[t]{0.5\columnwidth}%
\begin{center}
\subfloat[\label{fig:nu_mu}]{\begin{centering}
\includegraphics[scale=0.4]{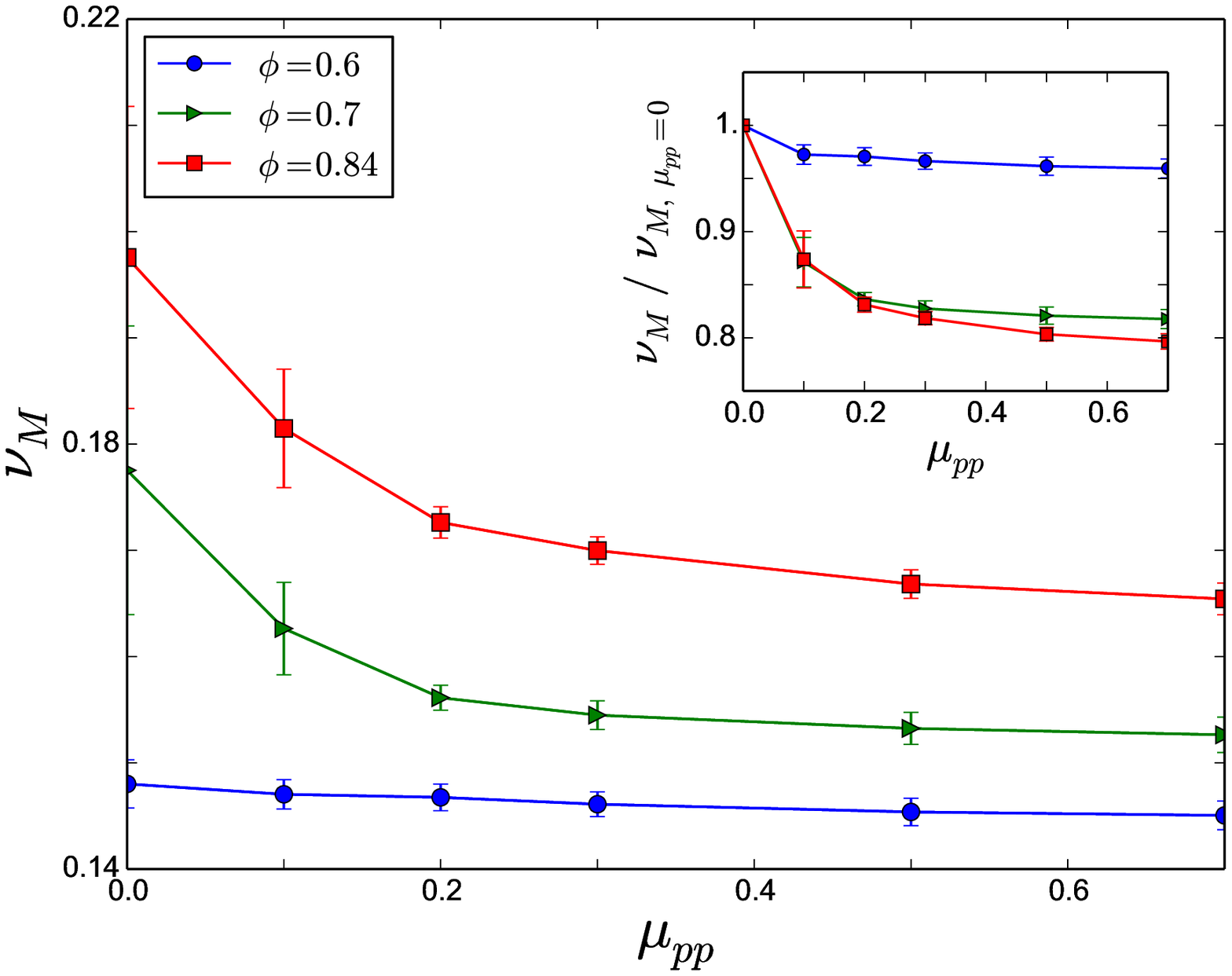}
\par\end{centering}

}

\par\end{center}%
\end{minipage}
\par\end{centering}

\caption{(a) Influence of the friction coefficient between pentagons on the macroscopic
Young's modulus for different solid fraction values. Macroscopic Young's modulus for each solid fraction has been normalized by its value for $\mu_{pp}=0$ to highlight the influence of the friction coefficient. This effect is
small although it increases with the solid fraction. Insert: Young's modulus for $\mu_{pp}=0$ versus the solid fraction. It increases as explained in section \ref{sub:Influence-of-solid}. (b) Macroscopic Poisson's ratio $\nu_{M}$ for the same values of $\phi$ and $\mu_{pp}$ than (a). Insert : $\nu_{M}$ normalized by $\nu_{M},\,\mu_{pp}=0$.} \label{fig:em_mu_nu}

\end{figure}

\par\end{center}
\begin{center}
\begin{figure}[htvp]
\begin{centering}
\includegraphics[scale=0.5]{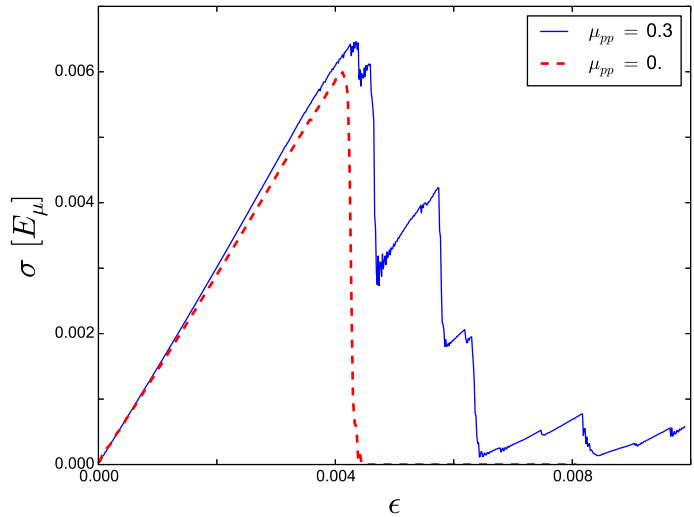} 
\par\end{centering}

\caption{Stress/strain curves for contact friction values $\mu_{pp}=0$ (red dots)
and $\mu_{pp}=0.3$ (blue) for the same sample of solid fraction $\phi=0.84$. The different post-peak behaviors are
clearly seen. \label{fig:stress-strain_mu}}
\end{figure}

\par\end{center}

Although the effect of the contact friction coefficient is rather
small in the elastic part, it has a strong influence on the post peak
behavior (Fig. \ref{fig:stress-strain_mu}). Indeed, with a friction coefficient of $0.$, the stress
falls suddenly to zero after having reached the stress peak, which
means that the fracture propagates quickly through the sample.
Contrariwise, a different behavior is observed with a contact friction
coefficient of $0.3$. First, the fracture initiates and the stress
drops. Then, we observe a stress increase until it reaches a new peak
lower than the main one. This means that the energy dissipation due
to contact friction influences the fracture propagation across the
sample, and so its strength in the post-peak region.

\subsubsection{Influence of the ratio $R/r$ \label{sub:influence_R}}

To introduce resistance to relative rotation for cohesive interactions, two ($2D$) or three ($3D$) cohesion points are placed at a distance $R$ of the center of each particle. For $R\neq\,0$, a relative rotation of the pentagons will induce a reaction torque, which increases with $R$ and so we expect an increase of macroscopic stiffness of the modeled material. As described before, $R$ has been chosen as the average radius $r$ of the smallest pentagon of the pair, thus $R/r=1$ for all cohesive interactions. Figure \ref{fig:em_r} shows the measured macroscopic Young's modulus for values of the ratio $R/r$ ranging from $0$ (no resistance to rotation) to $2$. It can be clearly seen that the ratio $R/r$ has a strong effect on the macroscopic Young's modulus, up to an increase by $20\%$ in the studied range, which also indicates that relative rotations are not negligible even for a solid fraction of $0.84$ as presented in Fig. \ref{fig:em_r}. However, the effect seems to saturate for values of $R/r$ above $1$ in this case. The ratio $R/r$ is thus a simple model parameter allowing to control the resistance of rotation between cohesive pairs of particles, and thus its influence on macroscopic stiffness of the material.

\begin{center}
\begin{figure}[htvp]
\begin{centering}
\includegraphics[scale=0.4]{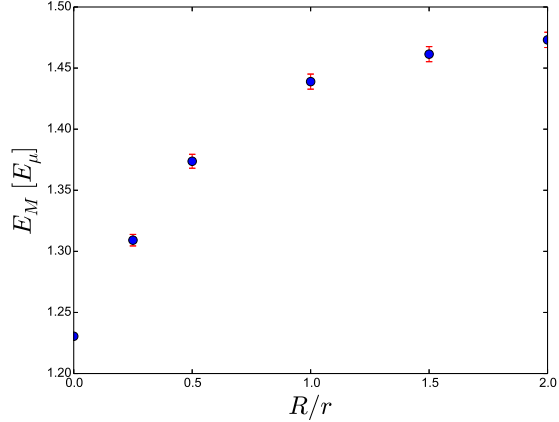} 
\par\end{centering}

\caption{Macroscopic Young's moduli $E_{M}$ for values of the ratio $R/r$ ranging from $0$ to $2$, for samples of solid fraction $\phi=0.84$.  \label{fig:em_r}}
\end{figure}
\par\end{center}

\subsubsection{Influence of cohesive interactions rupture threshold}

The macroscopic fracture of the material is only due to the rupture
of cohesive interactions between particles. Thus, the rupture threshold
$\sigma_{r}$ is expected to have a strong influence on the strength
of the modeled material. Simulations for different rupture threshold values
have been performed, and the corresponding stress/strain behaviors
are plotted in Fig. \ref{fig:stress_strain_sigr_a}. As expected,
the rupture threshold of cohesive interactions controls the peak stress
$\sigma_{peak}$ and peak strain $\epsilon_{peak}$ of the packing
and so the resistance of the material. However, it has no influence
on the slope of the elastic part of the stress-strain curves because no cohesive
interaction undergoes a stress as high as $\sigma_{r}$ during this
phase. The stress threshold also seems to change the post peak regime
by influencing the crack propagation, as can be seen in Fig. \ref{fig:stress_strain_sigr_b}.
The secondary peaks observed for simulations with $\sigma_{r}=5.10^{-3}$
are reduced with decreasing $\sigma_{r}$, and disappear for $\sigma_{r}=2.10^{-4}$.
Indeed when a crack initiates, weaker cohesive interactions will make
it easier to propagate across the sample. On the contrary, stronger
cohesive interactions stop the crack earlier and propagation of the
fracture will restart only when the strain energy becomes sufficient
again to break other cohesive interactions. In the case of very weak
cohesive interactions ($\sigma_{r}=2.10^{-4}$), the absence of secondary
peaks means that the first crack has enough energy to percolate across
the sample on its first try. This effect is superimposes with that of
contact friction described before to influence the propagation of
the rupture through the material.

\begin{center}
\begin{figure}[htvp]
\centering{}%
\begin{minipage}[t]{0.49\columnwidth}%
\begin{center}
\subfloat[\label{fig:stress_strain_sigr_a}]{\begin{centering}
\includegraphics[scale=0.4]{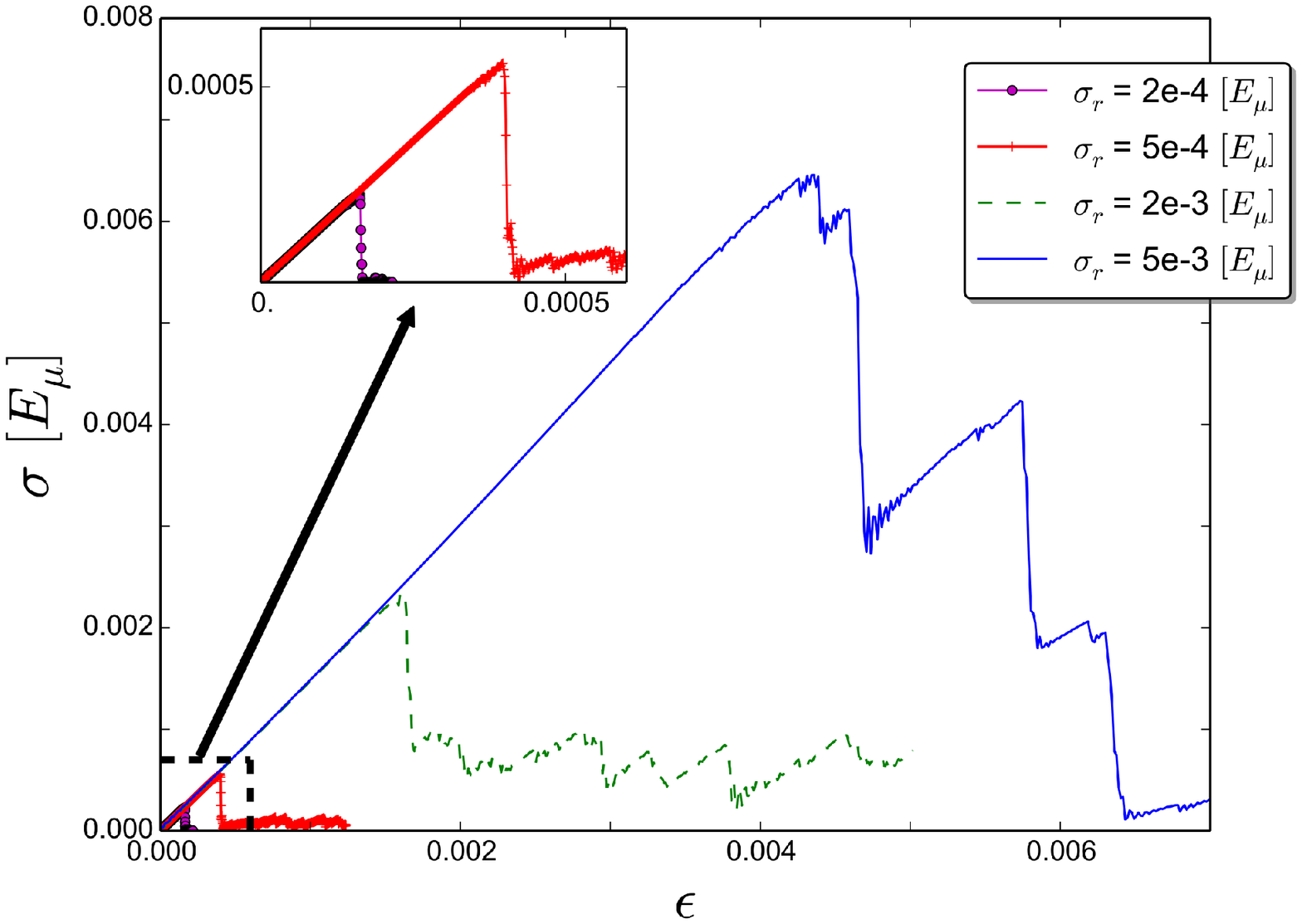}
\par\end{centering}

}
\par\end{center}%
\end{minipage}%
\begin{minipage}[t]{0.49\columnwidth}%
\begin{center}
\subfloat[\label{fig:stress_strain_sigr_b}]{\begin{centering}
\includegraphics[scale=0.4]{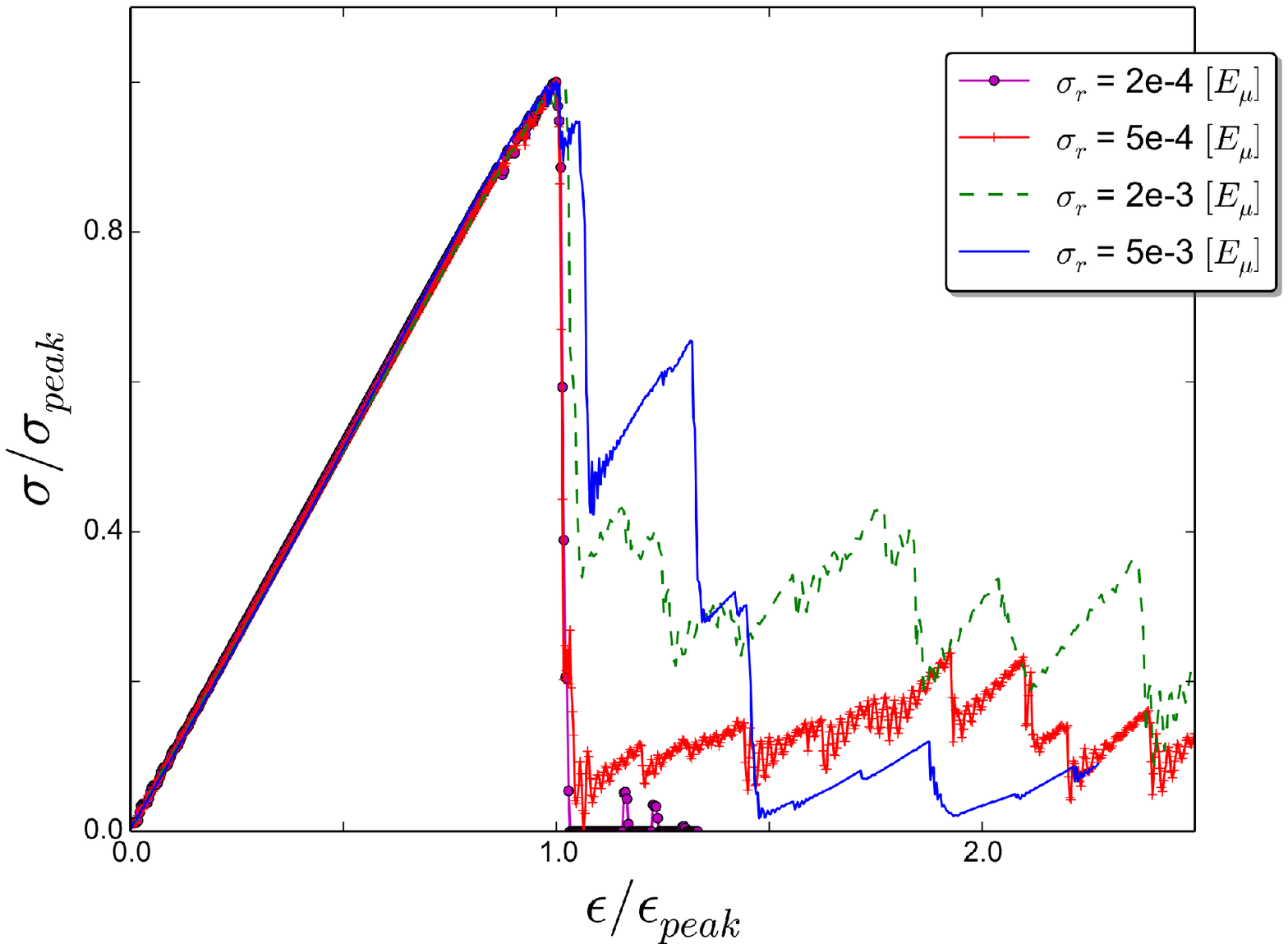}
\par\end{centering}

}
\par\end{center}%
\end{minipage}\caption{(a) Stress-strain behavior obtained for increasing cohesive interaction
rupture threshold. (b) Same as (a) but with the stress $\sigma$ and
axial strain $\epsilon$ normalized by values at peak ($\sigma_{peak}$,$\epsilon_{peak}$)
to see the effect of the stress threshold on post-peak behavior. \label{fig:stress-strain sigr}}
\end{figure}

\par\end{center}

\section{Conclusion}

We have presented a mixed model combining NSCD description for contact
behavior and cohesive interactions made of pairs of cohesion points
to mimic a cohesive phase lying between the particles of a rock piece.
The use of cohesion points allows to deal with particles of complex
shapes, and also to model cohesive materials of various solid fraction values up to 1.
The model was tested in $2D$, where an aggregate grain was modeled
as a polydisperse packing of polygons, bounded by cohesive interactions
in order to mimic the presence of cohesive bridges, as can be found
in real materials.

The model has proven to mimic the overall macroscopic behavior of
an aggregate grain subjected to axial compression. A linear elastic
behavior during which cohesive interactions remain intact is followed
by a sudden drop of stress on walls corresponding to the initiation
and propagation of macroscopic fractures inside the packing, characteristic
of a brittle material. Energy dissipation due to contact friction
has an important impact on the propagation of fractures across the
sample and induces progressive failure behavior. The microscopic elastic
limit allows to control the strength of the modeled material without
acting on the elastic slope of the response. As higher microscopic
elastic limit values require more energy to break cohesive interactions,
the fracture propagation is also influenced. The cohesive interactions
coordination number $z$ has an influence on both elastic and post-peak
behavior. Indeed, decreasing $z$ leads to a decrease of macroscopic
stiffness and has also an effect on breakage localization of cohesive
interactions. Finally, the solid fraction of the packing has been
studied and shows a direct effect on the macroscopic Young's modulus
and thus on the stiffness of the material.

The method used to model cohesion can be applied to any particle shape
in $2D$ and $3D$. Therefore, further studies will be carried out,
especially to quantify the influence of particle shapes for other
non regular polygons, closer to real material. The model will then
be tested for $3D$ simulations.

\section*{References}

%\bibliographystyle{bib_style}
%\bibliography{biblio}

\end{document}